\documentstyle[preprint,eqsecnum,aps]{revtex}
\draft
\tightenlines
\begin{document}

\title{{LOW-TEMPERATURE ASYMPTOTICS OF FREE ENERGY OF $3D$ 
ISING MODEL IN EXTERNAL MAGNETIC FIELD}}

\author{Martin S. Kochma\'nski} 
\address{Institute of Physics, Pedagogical University\\
T.Rejtana 16 A , 35--310 Rzesz\'ow, Poland\\
e-mail: mkochma@atena.univ.rzeszow.pl}
\maketitle

\begin{abstract}
The paper presents new method for calculating the low-temperature 
asymptotics of free energy of the $3D$ Ising model in external magnetic 
field $(H\neq 0)$. The results obtained are valid in the wide range of 
temperature and magnetic field values fulfilling the condition: 
$[1-\tanh(h/2)]\sim\varepsilon,$ for $\varepsilon\ll 1$, where $h=\beta H$, 
$\beta$ - the inverse temperature and $H$ - external magnetic field. 
For this purpose the method of transfer-matrix, and generalized Jordan-Wigner
transformations, in the form introduced by the author in $\cite{mkoch95}$,
are applied.
\end{abstract}

\widetext
\section{Formulation of the problem}
\label{sec: level1}
As is well known, till now an exact solution for the  
$2D$ Ising model in external magnetic field $(H\neq 0)$ was not found.
In the case of the $3D$ Ising model there does not exist an exact solution  
forvanishing magnetic field $(H=0)$, without even  mentioning the case with 
magnetic field. Despite great successes in investigations of the Ising model 
reachedusing the renormalization group method $\cite{wilson74}$, 
and other approximate methods $\cite{mccoy-wu73,ma76,sinaj80,curr91}$, the problem 
of calculation various asymptotics for the $2D$ and $3D$  Ising models
in external magnetic field  $(H\neq 0)$ is still of great importance.  
In the paper $\cite{koch97}$ we calculated low-temperature asymptotics for 
the $2D$  Ising model in external magnetic field $(H\neq 0)$, and free energy 
for this model in the limit  of asymptotically vanishing magnetic field. 
In this paper we shortly discuss the problem of calculation of 
the low-temperature asymptotics for free energy in the $3D$ Ising model 
in external magnetic field $(H\neq 0)$, following the approach and the ideas 
we introduced in the paper  $\cite{koch97}$.

Let us consider a cubic lattice built of $N$ rows, $M$ columns and 
$K$ planes, to vertices of which are assigned the numbers $\sigma_{nmk}$ 
from the two-entries set $\pm 1$. These quantities we will be calling here 
and everywhere below the Ising "spins." The multiindex $(nmk)$ numbers 
vertices of the lattice, with $n$ numbering rows, $m$ numbering columns, and 
$k$ numbering planes. The Ising model with the nearest neighbors ineteraction 
in external magnetic field is described by the Hamiltonian of the form:
\begin{equation}
{\cal H}=-\sum^{NMK}_{(n,m,k)=1}\left(J_{1}\sigma_{nmk}\sigma_{n+1,mk}+
J_{2}\sigma_{nmk}\sigma_{n,m+1,k}+J_{3}\sigma_{nmk}\sigma_{nm,k+1}+
H\sigma_{nmk}\right),
\end{equation}
taking into account anisotropy of the interaction between the nearest 
neighbors $(J_{1,2,3}>0)$, and the interaction of the spins $\sigma_{nmk}$  
with external magnetic field $H$, directed "up" $(\sigma_{nmk}=+1)$. 
The main problem consists of calculation of the statistical sum 
for the system: 
\begin{eqnarray*}
Z_3(h)=\sum_{\sigma_{111}=\pm 1} ... \sum_{\sigma_{NMK}=\pm 1}e^{-\beta\cal H}=
\end{eqnarray*}
\begin{equation}
\sum_{\{\sigma_{nmk}=\pm 1\}}\exp\left[\sum_{nmk}(K_1\sigma_{nmk}
\sigma_{n+1,mk}+K_2\sigma_{nmk}\sigma_{n,m+1,k}+K_3\sigma_{nmk}
\sigma_{nm,k+1}+h\sigma_{nmk})\right],
\end{equation}
where  $K_{1,2,3}={\beta}J_{1,2,3}, \;\;\; h={\beta}H, \;\;\;\beta=1/k_{B}T$. 
Typical boundary conditions for the variables $\sigma_{nmk}$ 
are the periodic ones. We  follow this standard assumption everywhere below. 
Let us note here that the statistical sum $(1.2)$ is symmetric with respect 
to the change $(h\rightarrow -h)$. 

In this letter we consider a limited version of the problem. Namely, the 
problem of calculation of the low-temperature asymptotics for free energy in 
the $3D$ Ising model in external magnetic field. More precisely, given the 
coupling constants $(J_{1,2,3}=const)$ and external magnetic field 
$(H=const)$,  we consider the region of temperatures satisfying the condition: 
$h\sim{\varepsilon}^{-1}, \;\;\;\;\; \varepsilon\ll 1$. To be more exact, we 
introduce a small parameter in the following way:
\begin{equation}
1 - \tanh(h/2)\sim\varepsilon, \;\;\;\;\; \varepsilon\ll 1 ,
\end{equation}
Then we consider the problem of calculation of free energy per one Ising spin 
in the thermodynamic limit, with exactness up to quantities of the order 
$\sim{\varepsilon}^2$ in expansions of the operators associated with 
interaction of spins as well among themselves as with the external field. 
(details of the approximation used will be presented below). 
In our opinion the problem formulated above is of reasonable 
importance, and, as far as is known to the author, it was not investigated in 
the existing literature. 

\section{Partition function}
\label{sec: level2}

Let us consider an auxillary $4D$  Ising model in external magnetic field $H$ 
on simple $4D$ lattice $(N\times M\times K\times L)$. We write the Hamiltonian
for the $4D$ Ising model with the nearest neighbor interaction in the form:
\begin{equation}
{\cal H}=-\sum_{n,m,k,l}\left(J_{1}\sigma_{nmkl}\sigma_{n+1,mkl}+
J_{2}\sigma_{nmkl}\sigma_{n,m+1,kl}+J_{3}\sigma_{nmkl}\sigma_{nm,k+1,l}+
J_{4}\sigma_{nmkl}\sigma_{nmk,l+1}+H\sigma_{nmkl}\right),
\end{equation}
taking into account anisotropy of the interaction between 
the nearest neighbors $(J_{1,2,3,4}>0)$, and interaction of the spins 
$\sigma_{nmkl}$ with external magnetic field $H$, directed "up" 
$(\sigma_{nmkl}=+1)$. Here  $(2.1)$ the multiindex $(nmkl)$ numbers the 
vertices of the $4D$ lattice, and the indices $(n,m,k,l)$ take on values from 
$1$ to $(N,M,K,L)$, respectively. As in the case of the $3D$ Ising model, 
we introduce periodic boundary conditions for the variables $\sigma_{nmkl}$. 
Then we write the partition function  $Z_4(h)$ in the form:
\begin{eqnarray*}
Z_4(h)=\sum_{\sigma_{1111}=\pm 1} ... \sum_{\sigma_{NMKL}=\pm 1}e^{-\beta\cal 
H}=\sum_{\{\sigma_{nmkl}=\pm 1\}}\exp\left[\sum_{nmkl}(K_1\sigma_{nmkl}
\sigma_{n+1,mkl} + \right. 
\end{eqnarray*}
\begin{equation}
\left.K_2\sigma_{nmkl}\sigma_{n,m+1,kl} + K_3\sigma_{nmkl}\sigma_{nm,k+1,l} 
+ K_4\sigma_{nmkl}\sigma_{nmk,l+1} + h\sigma_{nmkl})\right],
\end{equation}
where the quantities $K_{i}$ and $h$ are defined as above $(1.2)$  
$\cite{baxter82,izyum87}$. The expression  $(2.2)$ we can write, using the 
well known method of transfer matrix, in the form of a trace from the $L$-th 
power of the operator $\hat{T}$:
\begin{equation}
Z_4(h)=Tr(\hat{T})^L, \;\;\;\; \hat{T}=T_4T_h^{1/2}T_3T_2T_1T_h^{1/2},
\end{equation}
where the operators $T_{1,2,3,4,h}$ are defined by the formulas:
\begin{equation}
T_1=\exp\left(K_1\sum_{nmk}\tau^{z}_{nmk}\tau^{z}_{n+1,mk}\right), \;\;\;\;
T_2=\exp\left(K_2\sum_{nmk}\tau^{z}_{nmk}\tau^{z}_{n,m+1,k}\right),
\end{equation}
\begin{equation}
T_3=\exp\left(K_3\sum_{nmk}\tau^{z}_{nmk}\tau^{z}_{nm,k+1}\right), \;\;\;\;
T_4=(2\sinh 2K_4)^{NMK/2}\exp\left(K^{*}_{4}\sum_{nmk}\tau^{x}_{nmk}\right),
\end{equation}
\begin{equation}
T_h=\exp\left(h\sum_{nmk}\tau^{z}_{nmk}\right),
\end{equation}
and the quantities $K_4$ i $K^{*}_4$ are coupled by the following relations:
\begin{equation}
\tanh(K_{4})=\exp(-2K_{4}^{*}), \;\;\; or \;\;\; \sinh2K_{4}\sinh2K_{4}^{*}=1.
\end{equation}
The Pauli spin matrices $\tau^{x,y,z}_{nmk}$ commute for  $(nmk)\neq (n'm'k')$, 
and for given $(nmk)$ these matrices satisfy the usual relations 
$\cite{huang63}$. It is easy to see that the matrices $T_{1,2,3,h}$ commute 
among themselves, but do not commute with the matrix $T_4$. In the case in 
which one of the quantities $K_i=0,\;\; (i=1,2,3)$, we get obviously the 
known expressions describing the  $3D$ Ising model on a simple cubic lattice. 
Namely, the transition to the $3D$ Ising model with respect 
to the coupling constants  $K_1$,  $K_2$, or $K_3$ is realized by taking 
$(K_1=0)$, or $(K_2=0)$, or $(K_3=0)$, and removing summation over $n$, 
$(N=1)$, or over $m$, $(M=1)$, or over $k$, $(K=1)$, respectively. 
As a result we get the standard expressions $\cite{baxter82}$ for the $3D$ 
Ising model in external magnetic field. In the process the operators 
$T_i, \;\;(i=1,2,3)$ in every one of the cases are identically equal to the 
unit operator $(T_i\equiv \hat{1})$. A bit different situation appears in the 
case of the transition to the  $3D$ Ising model with respect to the coupling 
constant $K_4$. In this case we take $(K_4=0,\;\; L=1)$, i.e. we remove 
summation over $l$. In consequence we get the following expression for the 
operator $T_4$, $(2.5)$:
\begin{equation}
T^{*}_{4}\equiv T_4(K_4=0)=\prod_{nmk}(1+\tau^{x}_{nmk}) ,
\end{equation}
where we used the relation  $(2.7)$. Then, after transition to the limit  
$(K_4=0, \;\; L=1)$ in $(2.3)$, we can write the following expression for the 
partition function for the $3D$ Ising model: 
\begin{equation}
Z_3(h)=Tr(T_4^*T_h^{1/2}T_3T_2T_1T_h^{1/2}),
\end{equation}
where the matrices $T_i$ are defined as above $(2.4-6,8)$. Now we pass to the 
fermionic representation. For this aim one should write the matrices $T_i$ 
in terms of the Pauli operators $\tau_{nmk}^{\pm}$, $\cite{izyum87}$:
\begin{equation}
\tau^{\pm}_{nmk}=\frac{1}{2}(\tau^{z}_{nmk}\pm i\tau^{y}_{nmk}),
\end{equation}
which satisfy anticommutation relations for one vertex, and which commute for 
different vertices. 
 
As the next step one should pass from the representation by  Pauli  operators 
$(2.10)$ to the representation by Fermi creation and annihilation operators 
$\cite{mkoch95}$. In the paper $\cite{mkoch95}$ were  introduced appropriate 
transformations (generalized transformations of the Jordan-Wigner type), 
enabling the transition to the fermionic representation:
\begin{eqnarray}
\tau^+_{nmk}=\exp \left[ i\pi\left(\sum^{N}_{s=1}
\sum^{M}_{p=1}\sum^{k-1}_{q=1}\alpha^{+}_{spq}\alpha_{spq}+
\sum^{N}_{s=1}\sum^{m-1}_{p=1}\alpha^+_{spk}\alpha_{spk}+
\sum^{n-1}_{s=1}\alpha^+_{smk}\alpha_{smk}\right)\right]\alpha^{+}_{nmk}
\nonumber\\
\tau^+_{nmk}=\exp \left[ i\pi\left(\sum^{N}_{s=1}
\sum^{M}_{p=1}\sum^{k-1}_{q=1}\beta^{+}_{spq}\beta_{spq}+
\sum^{ n-1}_{s=1}\sum^{M}_{p=1}\beta^+_{spk}\beta_{spk}+
\sum^{m-1}_{p=1}\beta^+_{npk}\beta_{npk}\right)\right]\beta^+_{nmk}
\nonumber\\
\tau^+_{nmk}=\exp \left[ i\pi\left(\sum^{N}_{s=1}
\sum^{m-1}_{p=1}\sum^{K}_{q=1}\gamma^{+}_{spq}\gamma_{spq}+
\sum^{N}_{s=1}\sum^{k-1}_{q=1}\gamma^+_{smq}\gamma_{smq}+
\sum^{n-1}_{s=1}\gamma^+_{smk}\gamma_{smk}\right)\right]\gamma^+_{nmk}
\nonumber\\
\tau^+_{nmk}=\exp \left[ i\pi\left(\sum^{N}_{s=1}
\sum^{m-1}_{p=1}\sum^{K}_{q=1}\eta^{+}_{spq}\eta_{spq}+
\sum^{n-1}_{s=1}\sum^{K}_{q=1}\eta^+_{smq}\eta_{smq}+
\sum^{k-1}_{q=1}\eta^+_{nmq}\eta_{nmq}\right)\right]\eta^+_{nmk}
\nonumber\\
\tau^+_{nmk}=\exp \left[ i\pi\left(\sum^{n-1}_{s=1}
\sum^{M}_{p=1}\sum^{K}_{q=1}\omega^{+}_{spq}\omega_{spq}+
\sum^{M}_{p=1}\sum^{k-1}_{q=1}\omega^+_{npq}\omega_{npq}+
\sum^{m-1}_{p=1}\omega^+_{npk}\omega_{npk}\right)\right]\omega^+_{nmk}
\nonumber\\
\tau^+_{nmk}=\exp \left[ i\pi\left(\sum^{n-1}_{s=1}
\sum^{M}_{p=1}\sum^{K}_{q=1}\theta^{+}_{spq}\theta_{spq}+
\sum^{m-1}_{p=1}\sum^{K}_{q=1}\theta^+_{npq}\theta_{npq}+
\sum^{k-1}_{q=1}\theta^+_{nmq}\theta_{nmq}\right)\right]\theta^+_{nmk}
\end{eqnarray}
and analogously for the operators $\tau^{-}_{nmk}$. In the paper 
$\cite{mkoch95}$ we obtained  formulas for relations between various 
Fermi operators, and commutation relations for them. Further in this paper we 
will use the fact that the following equality of local occupation numbers is 
valid:
\begin{eqnarray}
\tau^+_{nmk}\tau^-_{nmk}&=&\alpha^+_{nmk}\alpha_{nmk}=\beta^+_{nmk}\beta_{nmk}=
\gamma^+_{nmk}\gamma_{nmk}=\eta^+_{nmk}\eta_{nmk}=
\omega^+_{nmk}\omega_{nmk}=\theta^+_{nmk}\theta_{nmk}.
\end{eqnarray}
Then, applying the expressions  $(2.10-12)$ and considerations from the paper 
$\cite{koch97}$, we can write the partition function $(2.9)$ in the form: 
\begin{equation}
Z_3(h)=(2\cosh^2h/2)^{NMK}<0|T^*|0>=A<0|U+{\mu}^2CUD|0>, \;\; 
U\equiv T_h^lT_3T_2T_1T_h^r, 
\end{equation}
where $A=(2\cosh^2h/2)^{NMK}$ and $\mu=\tanh(h/2)$, and the operators 
$T_{1,2,3}$, $T^{l,r}_h$ and $C,D$ are of the form: 
\begin{eqnarray}
T_1=\exp\left[K_{1}\sum_{n,m,k=1}^{N,M,K}(\alpha^{+}_{nmk}-\alpha_{nmk})
(\alpha^{+}_{n+1,mk}+\alpha_{n+1,mk})\right] ,  \nonumber\\
T_2=\exp\left[K_{2}\sum_{n,m,k=1}^{N,M,K}(\beta_{nmk}^{+}-\beta_{nmk})
(\beta^{+}_{n,m+1,k}+\beta_{n,m+1,k})\right] ,  \nonumber\\
T_3=\exp\left[K_{3}\sum_{n,m,k=1}^{N,M,K}(\theta_{nmk}^{+}-\theta_{nmk})
(\theta^{+}_{nm,k+1}+\theta_{nm,k+1})\right] ,
\end{eqnarray}
and
\begin{eqnarray}
T^r_h=\exp\!\left\{\!\mu^2\left[\sum_{nmk}\sum^{N-n}_{s=1}
\alpha^{+}_{nmk}\alpha^{+}_{n+s,mk}+\sum_{nn'mk}\sum^{M-m}_{t=1}
\alpha^{+}_{nmk}\alpha^{+}_{n',m+t,k}+\sum_{nn'mm'k}\sum^{K-k}_{l=1}
\alpha^+_{nmk}\alpha^+_{n'm',k+l}\right]\!\right\},\nonumber\\
T^l_h=\exp\!\left\{\!\mu^2\left[\sum_{nmk}\sum^{K-k}_{l=1}
\theta_{nm,k+l}\theta_{nmk}+\sum_{nmkk'}\sum^{M-m}_{t=1}
\theta_{n,m+t,k}\theta_{nmk'}+\sum_{nmm'kk'}\sum^{N-n}_{s=1}
\theta_{n+s,mk}\theta_{nm'k'}\right]\!\right\}, 
\end{eqnarray}
\begin{eqnarray*}
C=\sum_{nmk}\theta_{nmk}, \;\;\;\;\;\;\;\;\; D=\sum_{nmk}\alpha^+_{nmk}
\end{eqnarray*} 
Here and below $\sum_{n,m,...}$ means summation over the complete set of 
indices $(n=1,... N; \;\; m=1,... M; \;\;etc.)$. It is obvious that the 
operator  $\hat{G}$:
\begin{equation}
\hat{G}=(-1)^{\hat{S}}, \;\;\;\;\;\;\;\; \hat{S}=\sum_{nmk}\alpha^+_{nmk}\alpha_{nmk},
\end{equation}
where $\hat{S}$ is the operator of the total number of particles, commutes 
with the operator $T^*$, $(2.13)$. Therefore, we can divide all states of the 
operator $T^*$ into states with even $(\lambda_{\hat{G}}=+1)$ or odd number 
of particles $(\lambda_{\hat{G}}=-1)$ with respect to the operator $\hat{G}$, 
$(2.16)$. The form of the operators $T_{1,2,3}$ does not change during the 
course, only the boundary conditions for the operators $(\alpha_{nmk}, ...)$ 
do. In the case of even states $(\lambda_{\hat{G}}=+1)$ antiperiodic boundary 
conditions , and in the case of odd states periodic ones, are chosen 
$\cite{koch97}$. 

The next step is transition to the momentum representation:
\begin{eqnarray*}
\alpha^+_{nmk}=\frac{exp(i\pi/4)}{(NMK)^{1/2}}\sum_{qp\nu}e^{-i(nq+mp+k\nu)}
\xi^+_{qp\nu}, \;\;\;\;\; \beta^+_{nmk}\rightarrow\eta^+_{qp\nu}, \;\;\;\;\;\;
\theta^+_{nmk}\rightarrow\zeta^+_{qp\nu},
\end{eqnarray*}
and introduction for fixed  $(qp\nu)$ corresponding bases for $\xi$-, 
$\eta$- and $\zeta$- Fermi creation and annihilation operators in the 
representation in terms of  occupation numbers in the finite-dimensional Fock 
space of dimension $2^8=256$). Then, after a series of transformations and 
calculations  we arrive at the following formula for the partition function   
$(2.13)$: 
\begin{equation}
Z^{+}_{3D}(h)=A\left(\prod_{0<{q,p,\nu}<\pi}A^4_1(q)\right)\left(\prod_
{0<{q,p,\nu}<\pi}A^4_3(\nu)\right)<0|T^*_3(h)T_2T^*_1(h)|0>,
\end{equation}
where the operators $T_1^*(h), \;\;T_2, \;\;T^*_3(h)$ are of the form
\begin{eqnarray}
T^*_1(h)=\exp\left[\sum_{0<q,p,\nu <\pi}B_1(q)(\xi^+_{-q-p-\nu}\xi^+_{qp\nu}+ 
\xi^+_{-q-p\nu}\xi^+_{qp-\nu}+ \xi^+_{-qp-\nu}\xi^+_{q-p\nu}+ 
\xi^+_{-qp\nu}\xi^+_{q-p-\nu})\right], \nonumber\\
T_2=\exp\left\{2K_2\sum_{0<q,p,\nu <\pi}[\cos p(\eta^+_{qp\nu}\eta_{qp\nu} + 
...)+\sin p(\eta^+_{-q-p-\nu}\eta^+_{qp\nu} +
... + \eta_{qp\nu}\eta_{-q-p-\nu} + ...]\right\}, \nonumber\\
T^*_3(h)=\exp\left[\sum_{0<q,p,\nu <\pi}B_3(\nu)(\zeta_{qp\nu}\zeta_{-q-p-\nu} + 
\zeta_{-qp\nu}\zeta_{q-p-\nu} + \zeta_{q-p\nu}\zeta_{-qp-\nu} + 
\zeta_{-q-p\nu}\zeta_{qp-\nu})\right],
\end{eqnarray}
and  $A_1(q,h), ...$ are defined by the expressions: 
\begin{eqnarray}
A_1(q,h)=\cosh 2K_1-\sinh 2K_1\cos q+\alpha(h,q)\sinh 2K_1\sin q, \nonumber\\
A_3(\nu,h)=\cosh 2K_3-\sinh 2K_3\cos\nu+\alpha(h,\nu)\sinh 2K_3\sin\nu, \nonumber\\
B_1(q,h)=\frac{\alpha(h,q)[\cosh 2K_1+\sinh 2K_1\cos q]+\sinh 2K_1\sin q}
{A_1(q,h)},\nonumber\\
B_3(\nu,h)=\frac{\alpha(h,\nu)[\cosh 2K_3+\sinh 2K_3\cos\nu]+\sinh 2K_3\sin\nu}
{A_3(\nu,h)},\nonumber\\
\alpha(h,q)=\tanh^2(h/2)\frac{1+\cos q}{\sin q},\;\;\;\; 
\alpha(h,\nu)=\tanh^2(h/2)\frac{1+\cos\nu}{\sin\nu}.
\end{eqnarray}
In the formula for  $Z^+_{3D}(h)$  the sign $(+)$ 
means that we consider the case of even states 
$(\lambda_{\hat{G}}=+1)$ with respect to the operator $\hat{G}$, $(2.16)$. 
It is obvious that for $h=0$ we arrive at the $3D$ Ising model
in  vanishing magnetic field. Then, for $K_1=0$ (or $K_2=0,\;\; or\;\; K_3=0$) 
the expression $(2.17)$ for the statistical sum describes the $2D$ Ising model
in external magnetic field $\cite{koch97}$.

\section{{ Solution}}

Let us consider calculation of free energy per one Ising spin in external 
magnetoc field in the approximation described shortly in the introduction. 
For this aim let us consider the operators $T^*_1(h)$ and $T^*_3(h)$ in the 
"coordinate" representation:
\begin{eqnarray}
T^*_1(h)=\exp\left[\sum_{nmk}\sum^{N-n}_{s=1}a(s)
\alpha^{+}_{nmk}\alpha^{+}_{n+s,mk}\right],\nonumber\\
T^*_3(h)=\exp\left[\sum_{nmk}\sum^{K-k}_{l=1}c(l)
\theta_{nm,k+l}\theta_{nmk}\right],
\end{eqnarray}
where the "weights" $a(s)$ and $c(l)$ are defined by the formulas:
\begin{eqnarray}
a(s)=\frac{1}{N}\sum_{0<{q}<\pi}2B_1(q)\sin(sq)=
{z^*_1}^s+\tanh^2h^*_1\frac{1-{z^*_1}^s}{(1-z^*_1)^2},\;\;\;\;s=1,2,3,...\nonumber\\
c(l)=\frac{1}{K}\sum_{0<{\nu}<\pi}2B_3(\nu)\sin(l\nu)=
{z^*_3}^l+\tanh^2h^*_3\frac{1-{z^*_3}^l}{(1-z^*_3)^2},\;\;\;\;l=1,2,3, ...
\end{eqnarray}
We introduced renormalized quantities $(K^*_{1,3},\;\;h^*_{1,3})$  defined as 
follows :
\begin{eqnarray}
\sinh2K^*_{1,3}=\beta_{1,3}[\sinh2K_{1,3}(1-\tanh^2(h/2)],\nonumber \\
\cosh(2K^*_{1,3})=\beta_{1,3}[\cosh2K_{1,3}+\tanh^2(h/2)\sinh2K_{1,3}],
\nonumber \\
\beta_{1,3}=[1+2\tanh^2(h/2)\sinh2K_{1,3}e^{2K_{1,3}}]^{-1/2}, \;\;\; 
\tanh^2h^*_{1,3}=\tanh^2(h/2)\frac{\beta_{1,3}\exp(2K_{1,3})}{\cosh^2K^*_{1,3}},
\end{eqnarray}
These formulas are valid for $(K_{1,3}\geq 0)$. As in the case of the $2D$  
Ising model $\cite{koch97,999r.97}$, also in this case one can introduce a 
diagrammatic  representation for the vacuum matrix element 
$S\equiv<0|T^*_3(h)T_2T^*_1(h)|0>$. Computation of the vacuum matrix element 
$S$,  which enters the formula $(2.17)$ for $Z^+_{3D}(h)$ in general case,
where the "weights"  $(3.2)$ are arbitrary is, at least at present, 
impossible.  Nevertheless, there exists a special case in which we can 
calculate the quantity  $S$  in the $3D$ case. Namely, this is the case where 
the "weights" $(3.2)$ are independent of $l$ and $s$. In this case one should,  
as in the $2D$ case $\cite{koch96}$, put the parameters $K_{1,3}$  equal zero 
$(K_{1,3}=0)$ in the formula $(2.13)$, and then express the operators
$T^{l,r}_h$ in terms of the Fermi  $\beta$-operators $(2.11)$ of creation and 
annihilation, with the goal to calculate $S$. After transition to the 
momentum representation, one should calculate the vacuum matrix element 
$S^*(y_1,y_3,z_2)$:
\begin{eqnarray*}
S^*(y_1,y_3,z_2)\equiv<0|T^l(y_3)T_2T^r(y_1)|0>,
\;\;\;\;y_{1,3}\equiv\tanh^2h_{1,3},
\end{eqnarray*}
where $z_2=\tanh K_2$ becomes trivial.(Here we introduced  the following 
change of notation: $h/2\rightarrow h_1$ - for the operator $T^r_h$, and 
$h/2\rightarrow h_3$ - for the operator $T^l_h$). We can write the result for 
$S^*(y_1,y_3,z_2)$ in the following form:
\begin{eqnarray}
S^*(y_1,y_3,z_2)=(2\cosh^2K_2)^\frac{NMK}{2}\prod_{0<qp\nu<\pi}\left[(1-2z_2\cos p 
+ z^2_2)(1-cos p)+ 2z_2(y_1+y_3)\sin^2p + \right.\nonumber\\
\left.y_1y_3(1+2z_2\cos p + z^2_2)(1 + \cos p)\right]^4 .
\end{eqnarray}
This result can be used further to calculate free energy in the approximation 
discussed above $(1.3)$. For this aim let us note that the conditions 
$[\tanh^2h^*_{1,3}/(1-z^*_{1,3})^2]\rightarrow 1$ are equivalent, 
accordingly to $(3.3)$, to the conditions $(\exp(-2K_{1,3})(1-\tanh^2{h/2})\rightarrow 0)$. 
It follows from this equation that for fixed $(J_{1,3}=const,\;\; H=const)$ 
these conditions are satisfied in the region of temperatures $T$, in which 
$(h/2)\sim{\varepsilon}^{-1}, \;\;\;\varepsilon\ll 1$. In this case we 
can use the result $(3.4)$. Namely, let us consider the formulas  $(2.19)$ 
for $B_{1,3}$, written in terms of the renormalized parameters 
$(h^*_{1,3},\;\;K^*_{1,3})$:
\begin{equation}
B_{1,3}=\frac{\tanh^2h^*_{1,3}\frac{\sin q(\nu)}{1-\cos q(\nu)}+2z_{1,3}^*\sin 
q(\nu)}{1-2z_{1,3}^*\cos q(\nu) +{z_{1,3}^*}^2},
\end{equation}
where $z^*_{1,3}=\tanh K^*_{1,3}$. Next, since the following equalities are 
satisfied: 
\begin{eqnarray*}
\frac{z_{1,3}^*}{1+{z_{1,3}^*}^2}=\frac{z_{1,3}(1-\tanh^2{h/2})}{1+
2z_{1,3}\tanh^2{h/2}+z^2_{1,3}},
\end{eqnarray*}
then, if we introduce a small parameter  $[1-\tanh(h/2)]\sim\varepsilon, \;\;\; (\varepsilon\ll 1)$, 
and expand  $B_{1,3}$ into a series in powers of  $\varepsilon$  
$(z^*_{1,3}\sim\varepsilon)$, we obtain 
\begin{eqnarray*}
B_{1,3}=\frac{(\tanh^2h^*_{1,3}+2z^*_{1,3})\sin q(\nu)}{1-\cos q(\nu)}+
\sim{\varepsilon}^2
\end{eqnarray*}
This formula gives the following expressions for the "weights"$a(s)$ and 
$c(l)$, $(3.2)$ in this approximation:
\begin{equation}
a(s)= \tanh^2h^*_1+2z^*_1,\;\;\;\;\; c(l)=\tanh^2h^*_3+2z^*_3  ,
\end{equation}
with exactness of the order of smallness $\sim{\varepsilon}^2$. As a result 
in this approximation the "weights" $a(s), \;\;c(l)$ do not depend on  
$(s,l)$. Finally, if we substitute to the expression $(3.4)$ for $S^*(y_1,y_3,z_2)$ 
the parameters $y_1\rightarrow a(s)$ and $y_3\rightarrow c(l)$, $(3.6)$, we 
arrive at the following formula for free energy on one  Ising spin 
$F_{3D}(h)$  in the thermodynamic limit:
\begin{eqnarray}
-\beta F_{3D}(h)\asymp\ln(2^{3/2}\cosh K^*_1\cosh K_2\cosh K^*_3\cosh^2{h/2})+
\frac{1}{2\pi}\int^{\pi}_0\ln\left[(1-2z_2\cos p+z^2_2)\times\right.\nonumber\\
\left.(1-\cos p)+2z_2(\tanh^2h^*_1+\tanh^2h^*_3+2z^*_1+2z^*_3)\sin^2p+(\tanh^2h^*_1+
2z^*_1)(\tanh^2h^*_3+\right. \nonumber\\
\left.2z^*_3)(1 + 2z_2\cos p + z^2_2)(1 + \cos p)\right]dp,
\end{eqnarray} 
where $\beta=1/k_BT$, and $z_2=\tanh K_2$, and $h^*_{1,3}$ and $K^*_{1,3}$ 
are coupled with  $h$ and $K_{1,3}$ by the relations $(3.3)$. One can show 
that, as it was done for the $1D$ and $2D$ Ising models $\cite{koch97,koch96}$, 
in the case of the states odd $(\lambda_{\hat{G}}=-1)$  with respect to the 
operator $\hat{G}$, $(2.16)$, the formula for  $F_{3D}(h)$ is described in 
the thermodynamic limit by $(3.7)$. Let us note that the asymptotics  $(3.7)$ 
obtained above can be applied also in the case  of rather strong magnetic 
fields  $(H)$, as far as it satisfies the condition  
$(1-\tanh h)\sim\varepsilon, \;\;\; \varepsilon\ll 1, \;\; (T=const)$.

\section{Final remarks}

The result derived above$(3.7)$ can be applied to the analysis of equilibrium 
thermodynamics of the three dimensional Ising magnetic, lattice gas, and also 
three dimensional models of binary alloys $\cite{ziman79,thompson88}$ 
in the region of temperatures and magnetic fields $(1.3)$ derived above. Such 
analysis, as well as construction of appropriate phase diagramms for the 
models mentioned above is, in our opinion, of great interest. They deserve   
presentation in a separate publication. Therefore we deliberately do not 
compare here our result $(3.7)$ with the existing papers devoted to this 
problem. The other important feature of the presented method is the 
possibility of deriving expressions for the free energy of the $3D$ Ising 
model in the limiting case of the magnetic field  tending to zero 
$(H\rightarrow 0, \;\;\; N,M,K\rightarrow\infty)$ if we know exact solution 
for the $3D$ Ising model in the absence of external magnetic field 
$(H=0)$. This possibility results from equations $(3.2-3.3)$ describing 
renormalised interaction constants $K^*_{1,3}$, and corresponds, as was 
presented in the paper $\cite{koch97}$, to the results obtained by 
C.N.Yang $\cite{yang52}$ for the $2D$ Ising model. 

\acknowledgements

I am grateful to H. Makaruk, R. Owczarek  for their assistance in 
preparation of the final form of this paper.

\end{document}